\documentclass[prl,
                twocolumn,
                preprintnumbers,
                amsmath,
                amssymb,
                showpacs,
                superscriptaddress]{revtex4-1}
\usepackage{graphicx} 
\usepackage{dcolumn}  
\usepackage{bm}       
\usepackage{amsfonts}
\usepackage[colorlinks=true,linkcolor=blue,citecolor=red,urlcolor=blue]{hyperref}
\usepackage{color}
\newcommand{\eq}[1]{Eq.~(\ref{#1})}
\newcommand{\ful}{\mbox{C$_{\mbox{\scriptsize{60}}}$}}

\begin{document}

\title{Ubiquitous diffraction resonances in positronium formation from fullerenes}

\author{Paul-Antoine Hervieux}
\email[]{hervieux@unistra.fr}
\affiliation{%
Universit\'e de Strasbourg, CNRS, Institut de Physique et Chimie des Mat\'eriaux de Strasbourg, 67000 Strasbourg, France}

\author{Anzumaan R. Chakraborty}
\affiliation{%
Department of Natural Sciences, D.L.\ Hubbard Center for Innovation and Entrepreneurship,
Northwest Missouri State University, Maryville, Missouri 64468, USA}

\author{Himadri S. Chakraborty}
\email[]{himadri@nwmissouri.edu}
\affiliation{%
Department of Natural Sciences, D.L.\ Hubbard Center for Innovation and Entrepreneurship,
Northwest Missouri State University, Maryville, Missouri 64468, USA}

\date{\today}

\pacs{34.80.Lx, 36.10.Dr, 61.48.-c}


\begin{abstract}
Due to the dominant electron capture by positrons from the molecular wall and the spatial dephasing across the wall-width, a powerful diffraction effect universally underlies the positronium (Ps) formation from fullerenes. This results into trains of resonances in the Ps formation cross section as a function of the positron beam energy, producing uniform structures in recoil momenta in analogy with classical single-slit diffraction fringes in the configuration space. The prediction opens a hitherto unknown avenue of Ps spectroscopy with nanomaterials.
\end{abstract}

\maketitle

Following the impact of positrons with matter the formation of exotic electron-positron bound-pair, the positronium (Ps), is a vital process in nature. This channel accounts for as large as half of the positron scattering cross section from simple atoms and molecules~\cite{laricchia2008}, as well as an even higher success rate of Ps formation on surfaces and thin films~\cite{schultz1988}. Other than probing structure and reaction mechanism of matters, the Ps formation is a unique pathway to the electron-positron annihilation process~\cite{green2015,kauppila2004} with both astrophysical~\cite{prantzos2011} and applied~\cite{kavetskyy2014} interests. Possible production of Bose-Einstein condensate of Ps has also been predicted~\cite{shu2016,mor2014}, besides the importance of Ps in diagnosing porous materials~\cite{cassidy2008} as well as in probing bound-state QED effects~\cite{karshenboim2004}. Moreover, efficient Ps formation is the precursor of the production of dipositronium molecules~\cite{cassidy2007} and antihydrogen atoms~\cite{ferragut2010, mcconnell2016} required to study the effect of gravitational force on antimatter~\cite{crivelli2014, Perez2015}.

Theoretical investigations to calculate Ps formation cross sections from atomic hydrogen \cite{yamanaka2001,kamali2001,kar2000}, noble gases~\cite{sen2009,ghanbari2013,mceachran2013}, and alkali metals~\cite{gianturco1996,pandey2016} are aplenty. Calculations with molecular targets, although relatively limited, include the molecular hydrogen~\cite{biswas2002}, polyatomic molecules~\cite{sueoka2000}, and the water molecule~\cite{hervieux2006}. Simultaneously, precision experimental techniques to measure Ps formation signals have also been achieved by impinging positrons into varieties of materials, such as, atomic and molecular gases~\cite{garner1998,machacek2016}, molecular solids~\cite{eldrop1983}, liquids and polymers~\cite{wang1998}, zeolites~\cite{cabral-prieto2013}, metal surfaces and films~\cite{cooper2016}, metal-organic-frameworks~\cite{crivelli2014mof,jones2015}, and mesostructures~\cite{andersen2016}. Very recently high yields of laser assisted production of low-energy excited Ps is achieved in the interaction of cold-trapped positrons with Rydberg excited Cs atom~\cite{mcconnell2016}.

However, in spite of such broad landscape of Ps research, little attempt of Ps formation by implanting positrons in nanoparticles, in gas or solid phase, has so far been made, except for a single theoretical study using Na clusters~\cite{fojon2001}. On the other hand, straddling the line between atoms and condensed matters are clusters and nanostructures that not only have hybrid properties of the two extremes, but also exhibit outstanding behaviors with unusual spectroscopic effects~\cite{jaenkaelae2011}. Formation of a quasi-free electron gas within a finite nanoscopic region of well-defined boundary, as opposed to a longer-range, highly diffused Coulomb-type boundary characteristic of atoms and molecules, is a property of nanosystems which ensures predominant electron capture from localized regions in space. This may lead to diffraction in the capture amplitude, particularly at positron energies that cannot excite plasmon modes. The Ps formation from fullerenes can be singularly attractive to access this diffraction phenomenon due to fullerene's eminent symmetry and stability, and its previous track record of success in spectroscopic experiments~\cite{ruedel2002oscExp}. In this communication, we show that the Ps formation amplitude from the positron colliding with $\ful$ does include a strong diffraction effect, resulting in a system of peaks in the form of broad shape resonances, in the total ground and excited Ps formation cross section for state-selected captures.
\begin{figure}[h!]
\includegraphics[angle=0,width=8.5cm]{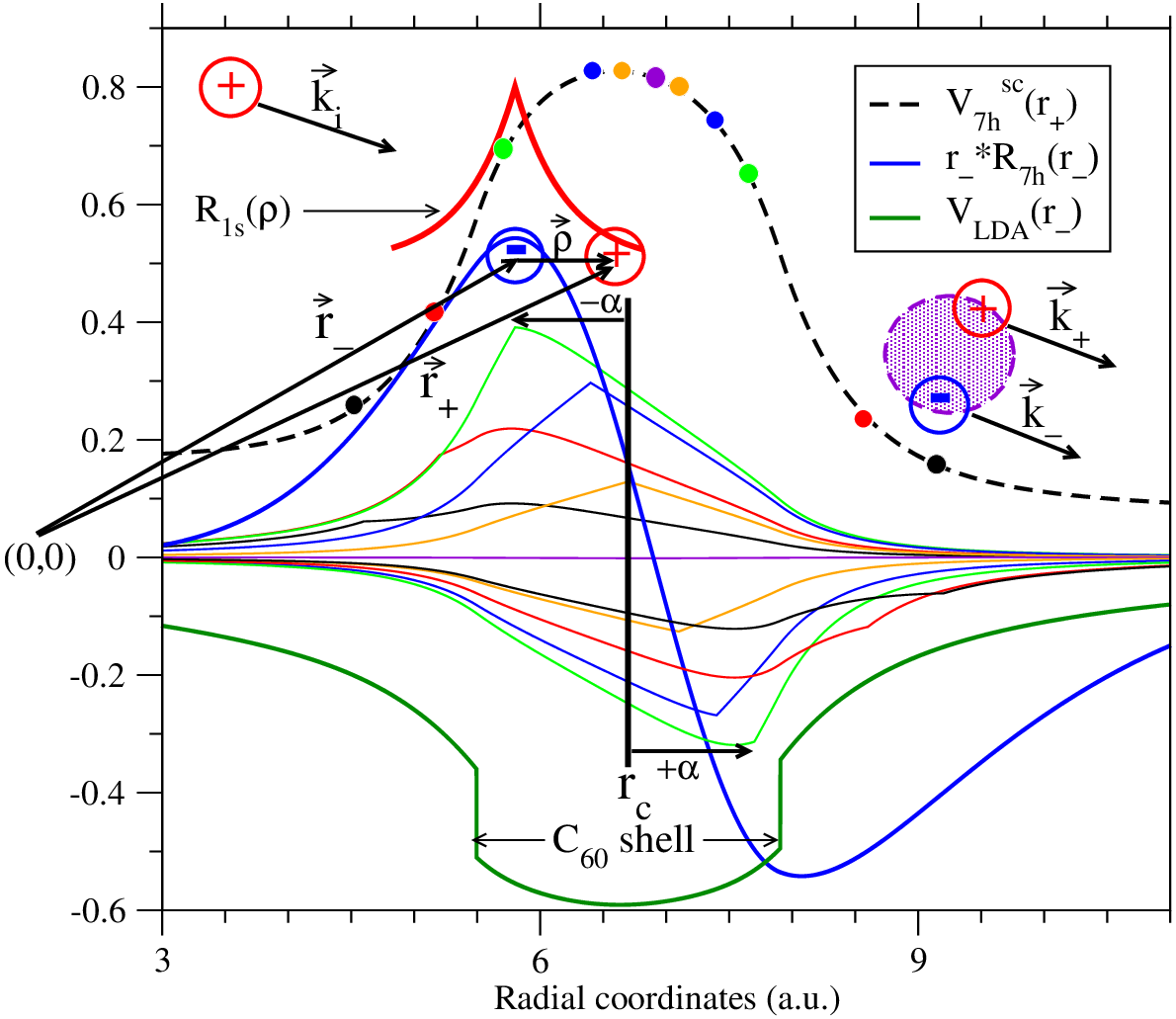}
	\caption{(Color online) The scattering potential $V_{2h}^{sc}$ of the positron from C$_{60}^+$ after a 7$h$ (HOMO) electron is captured, the radial HOMO wavefunction and the free $\ful$ ground state LDA radial potential identifying the shell width. Various position and momentum (for capture in the forward direction) vectors are schematically shown. A bunch of curves peaking around the $\ful$ radius $r_c$ represents products of scattering potential, radial HOMO and radial Ps($1s$) (shown) wavefunctions for the choices of electron positions marked by correspondingly color-coded dots. See text for the description of $\alpha$.}
	\label{figure1}
\end{figure}

The $\ful$ molecule is modeled by smearing sixty C$^{4+}$ ions in a spherical jellium shell, fixed in space, with an experimentally known $\ful$ mean radius $r_c$ = 6.7 a.u. and a width determined {\em ab\,initio}~\cite{madjet2008}. Inclusion of molecular orientations will have minimal effect on the result due to the $\ful$ symmetry~\cite{ciappina2007}. The delocalized system of total 240 valence electrons from sixty carbon atoms constructs the ground state in the Kohn-Sham local-density approximation (LDA)~\cite{madjet2008} improved by the gradient-corrected Leeuwen and Baerends exchange-correlation functional (LB94)~\cite{van1994exchange}. This produced bands of $\pi$ (one radial node) and $\sigma$ (nodeless) states with HOMO and \mbox{HOMO-1} to be of 7$h$ ($\ell_i=5$) and 6$g$ ($\ell_i=4$) $\pi$ character respectively -- a result known from the quantum chemical calculations~\cite{troullier1992} supported by direct and inverse photoemission spectra~\cite{weaver1991}, and from energy-resolved electron-momentum density measurements~\cite{vos1997}. The LDA radial ground state potential and the radial HOMO wavefunction are shown in Figure 1. Linear response type calculations using this ground state basis well explained measured photoemission response of $\ful$ at the plasmon excitation energies~\cite{madjet2008,scully2005volumeExp}. Similar calculations at higher energies also supported an effective fullerene width accessed in the experiment~\cite{ruedel2002oscExp}. Jellium-type modeling of multiwalled carbon nanotubes also had success in describing laser pump-probe measurements~\cite{zamkov2004}.

Consider an incoming positron of momentum $\vec{k}_i$ which captures an electron from a $\ful$ bound state $\phi_i(\vec{r}_-)$ to form a Ps state $\phi_f(\vec{\rho})$. As illustrated in Fig.\,1, the positron and electron position vectors, $\vec{r}_+$ and  $\vec{r}_-$ respectively, originate from the center of $\ful$ so that $\vec{\rho}=\vec{r}_- - \vec{r}_+$, and $\vec{k}_{+(-)}$ denote positron (electron) outgoing momenta in Ps that are equal, resulting $2\vec{k}_{+(-)}$ to be the momentum of Ps itself. Since we access energies above $\ful$ plasmon resonances, the many-body effect is not important, justifying the use of mean-field LDA wavefunctions and potentials in the framework of independent particle model. Therefore, the {\em prior} form of the Ps formation amplitude can be given in the continuum distorted-wave final-state (CDW-FS) approximation~\cite{fojon1996,fojon2001} as,
\begin{equation}\label{tot-amp}
T(\vec{k}_i) \sim \int d\vec{r}_- F_{\vec{k}_-}^{(-)^\ast}(\vec{r}_-) W(\vec{r}_-;\vec{k}_i) \phi_i(\vec{r}_-),
\end{equation}
in which
\begin{eqnarray}\label{pos-amp}
W(\vec{r}_-;\vec{k}_i) &=& \int d\vec{r}_+ F_{\vec{k}_+}^{(-)^\ast}(\vec{r}_+) \phi_f^\ast(\vec{\rho})\nonumber \\
                       && \times \left[V_i^{sc}(r_+) - \frac{1}{\rho}\right]F_{\vec{k}_i}^{(+)}(\vec{r}_+),
\end{eqnarray}
where the short-range positron-C$_{60}^+$ interaction, after the $i$-th electron is captured, {\em plus} the long-range $1/r_+$ interaction between them gives the positron scattering potential, $V^{sc}_i(r_+)$, from the C$_{60}^+$ ion; the LDA version of this potential for a HOMO electron capture is given in Fig.\,1. $1/\rho$ is the Coulomb interaction between the positron and the captured electron. For a given electron position, \eq{pos-amp} embodies the {\em snapshot} amplitude for the transition of an incoming positron to a Ps-bound outgoing positron. $W(\vec{r}_-;\vec{k}_i)$ thus provides the perturbation in \eq{tot-amp} for the capture of the bound electron into a moving Ps. We use \eq{tot-amp} {\em exactly} to compute our results, but consider simplifications to interpret them.

Integrations over $1/\rho$ will produce a steady contribution $S(\vec{k}_i)$ to the amplitude in energy which we ignore temporarily. This steady behavior may be justified by the steady Ps formation cross section of atomic hydrogen $1s$ capture into Ps($1s$) [Fig.\,2(a)] where $V_i^{sc}$ in \eq{pos-amp} is simply replaced by $1/r_+$. Also, to better explain our results, we approximate the three distorted Coulomb continuum waves, representable by confluent hypergeometric functions~\cite{fojon2001}, in Eqs.\, (\ref{tot-amp}) and (\ref{pos-amp}) as plane waves of the form $F_{\vec{k}}^{(-)}(\vec{r}) \sim \exp(-i\vec{k}\cdot\vec{r}$). These plane waves can expand in spherical harmonics:
\begin{equation}\label{plane-ex}
\exp(-i\vec{k}\cdot\vec{r}) \sim \sum_{\ell,m} i^l j_\ell(kr) Y_{\ell,m}(\hat{k}) Y_{\ell,m}^\ast(\hat{r}).
\end{equation}
With these and assuming the capture in an $n_fs$ Ps state $\phi_f(\vec{\rho}) \sim R_{n_fs}(\rho)Y_{0,0}(\hat{\rho})$, \eq{pos-amp} simplifies to
\begin{equation}\label{pos-amp2}
W(\vec{r}_-;\vec{k}_i) \sim \int dr_+r_+^2 j_0(qr_+) R_{n_fs}(\rho) V_i^{sc}(r_+) Y_{0,0}^\ast(\hat{q})
\end{equation}
where $\vec{q}=\vec{k}_+ - \vec{k}_i$ is the momentum transfer vector and \eq{pos-amp2} is isotropic in $\vec{q}$. (In writing \eq{pos-amp2}, we assume Ps formations in the forward direction, which was found to be the most dominant direction in both earlier~\cite{tang1992} and recent experiments~\cite{shipman2015}, so that $\hat{r}_-$ and $\hat{r}_+$ are identical and thus $\vec{\rho}$ is independent of $\hat{r}_+$.) Adopting the same simplification for $F_{\vec{k}_-}^{(-)^\ast}(\vec{r}_-)$ in \eq{tot-amp} to apply in \eq{plane-ex} and then combining with \eq{pos-amp2}, we can rewrite the full amplitude (\ref{tot-amp}) for a capture from $\ful$ $n_i\ell_i$ state as,
\begin{eqnarray}\label{tot-amp2}
T(\vec{k}_i) &\sim& S(\vec{k}_i) + \frac{1}{k_-q}\iint dr_-dr_+r_+ R_{n_fs}(\rho) V_i^{sc}(r_+) \nonumber \\
             &&\times [r_-R_{n_i\ell_i}(r_-)] \sin(k_-r_- - \ell_i\pi/2)\sin(qr_+) \nonumber \\
						 &&\times Y_{0,0}^\ast(\hat{q}) Y_{\ell_i,m_i}^\ast(\hat{k}_-),
\end{eqnarray}
where we used asymptotic forms of the spherical Bessel function $j_\ell(r)\sim \sin(r-\ell\pi/2)/r$. We will ignore the spherical harmonics in \eq{tot-amp2}, since the forward direction is considered.

Before moving further let us note the following in  the amplitude (\ref{tot-amp2}) in conjunction with Fig.\,1: (i) Large values of $V_i^{sc}$ only at the $\ful$ shell indicates the shell to be the window of appreciable repulsive interactions of the positron with $\ful$. (ii) The shape of the Ps radial wavefunction $R_{n_fs}(\rho)$ identifies the region of success of Ps formation as a function of electron-positron separation $\rho=|\vec{r}_--\vec{r}_+|$ where a Ps s-state has the maximum probability density at $r_-=r_+$. (iii) The radial wave function of $\ful$ $i$-th level $r_-R_{n_i\ell_i}$ ensures that electrons must be available for the Ps to form. Considering Ps($1s$) formation from the capture off $\ful$ HOMO level, we took eleven electron-positions in Fig.\,1 within the interaction window around $\ful$ radius ($r_c$) and plot the product $R_{1s}(\rho) V_{7h}^{sc}(r_+) [r_+ R_{7h}(r_+)]$ of the three quantities mentioned above; the $r_-$ values considered are shown by dots in the same color of that of the curve it generates. Since Ps will have a propensity to successfully form toward $\rho=0$, the curves show maxima and minima at various locations symmetrically on the left and right side, respectively, of $r_c$. This opposite orientation is because of the shape of $\ful$ $7h$ wavefunction which is asymmetric about its node. It is fair to approximate that the dominant behavior comes from the optimum positions at $r_p$. Thus, choosing $r_-=r_+=r_p$ in \eq{tot-amp2} these contributions indicate trains of peaks given by $\sin(k_-r_- - \ell_i\pi/2)\sin(qr_+) \sim \sin(Qr_p - \ell_i\pi/2)$ where the recoil momentum $Q=k_i-2k_\pm$ for the Ps formation in the forward direction. The amplitude $A(r_p)$ of the oscillation is proportional to the area ($r_+$ integral) under the corresponding bump. Hence, the remaining $r_-$ integral in \eq{tot-amp2} can approximate to a simpler form
\begin{equation}\label{tot-amp3}
T(\vec{k}_i) \sim S(\vec{k}_i) + \frac{1}{k_-q}\int dr_- A(r_p) \sin(Qr_p - \ell_i\pi/2).
\end{equation}
Noticing that the bumps appear on either side of $r_c$ at similar distances $\alpha$ (Fig.\,1), \eq{tot-amp3} can be rewritten as
\begin{eqnarray}\label{tot-amp4}
T(\vec{k}_i) &\sim& S(\vec{k}_i) + \frac{1}{2k_-q}\int dr_- A(r_p)[\sin\{Q(r_c-\alpha(r_p)) \nonumber \\
             &&- \ell_i\pi/2)\} -\sin\{Q(r_c+\alpha(r_p)) - \ell_i\pi/2)\}],
\end{eqnarray}
where the negative sign between the two terms justifies the maxima and minima about $r_c$ (Fig.\,1).
We finally obtain
\begin{eqnarray}\label{tot-amp5}
T(\vec{k}_i) &\sim& S(\vec{k}_i) - \frac{1}{k_-q}\cos(Qr_c - \ell_i\pi/2) \nonumber \\
             && \times \int dr_- A(r_p)\sin(Q\alpha(r_p)).
\end{eqnarray}

Obviously, the integral in the above equation spatially dephase the $\sin(Q\alpha)$ modulation, since $\alpha$ varies with $r_-$, retaining peaks in the amplitude only via $\cos(Qr_c - \ell_i\pi/2)$. Here is the essence of what is going on. Within about the fullerene width region, a strong constructive and destructive interference take place as a function of $Q$. When the odd integer multiple of the half-wavelength of {\em effective} continuum wave as a function of $Q$ fits the distance $r_p$, systems of peaks (bright-spots) in the energy domain are formed which subsequently results into a single {\em centroid} fringe pattern via a dephasing mechanism in the integration over the electron position as described above.
\begin{figure}[h!]
\vskip 0.5cm
\includegraphics[width=8cm]{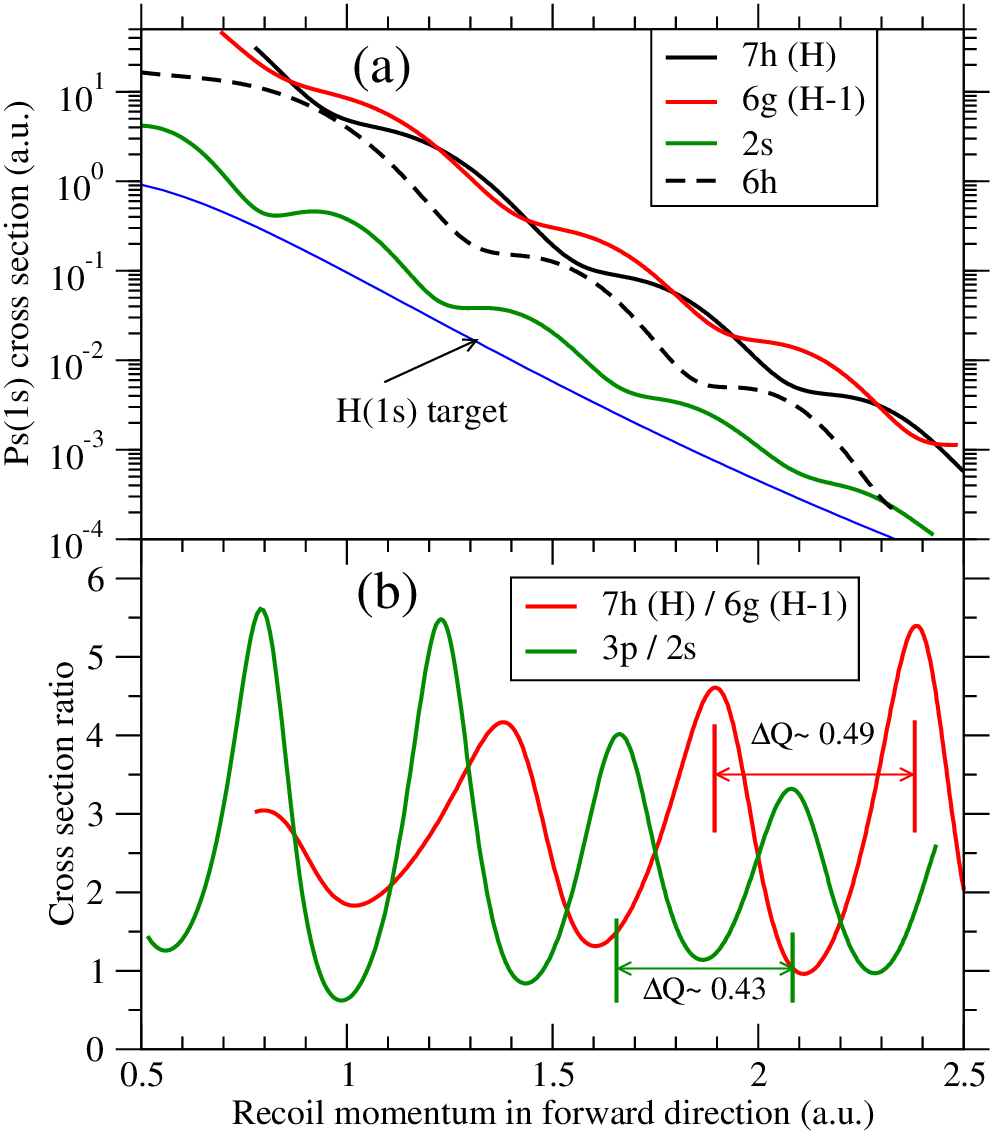}
	\caption{(Color online) (a) Ps($1s$) formation cross sections for captures from HOMO (H), HOMO-1 (H-1), $2s$ (bottom most $\pi$ level) and $6h$ (a $\sigma$ level) as a function of recoil momentum ($Q$) in the forward direction. The result of hydrogen $1s$ capture is also shown for comparisons. (b) The ratios of two pairs of these cross sections illustrate strong resonances. Separations ($\Delta Q$) between some resonances are marked.}
	\label{figure2}
\end{figure}

Since these diffraction peaks appear in the energy (momentum) domain, we call them diffraction resonances which must also show up in the Ps formation cross section proportional to the squared modulus of \eq{tot-amp5}. Further, this squaring operation modifies the functional form of the structures as $\cos(Qd_c - \ell_i\pi)$ in the cross section, where $d_c$ is the $\ful$ diameter. Broad resonances are seen in Fig.\,2(a) for the numerical Ps($1s$) formation cross section for the captures off four $\ful$ levels as a function of the forward-emission recoil momentum $Q$ corresponding to the electron excitation energy from roughly 50 eV (above the plasmon excitation) to 270 eV (below the $K$-shell of atomic carbon). For a direct comparison, also presented in Fig.\,2(a) is the steady result for the atomic hydrogen target showing no such resonances. The off-sets between the peaks for the HOMO versus HOMO-1 versus $2s$ (the bottom most level of $\ful$ $\pi$ band) results connect to the phase-shift $\ell_i\pi/2$ in \eq{tot-amp5} that depends on the angular symmetry of the initial state. The shape and strength of these resonances are best illustrated by considering the cross section ratios, shown in Fig.\,2(b), which neutralizes the non-resonant background decays, and therefore can be accessed in experiments by the Ps formation spectroscopy with better accuracy much freer of the experimental noise. It is important to note that we used plane wave descriptions in our analysis to interpret the key results. But the actual character of the resonances are more diverse than this simple account, as seen in Fig.\,2(b).

If the capture is from a $\ful$ $\sigma$ state, which has no radial node, we still also find \eq{tot-amp4} but with a positive sign between the two terms in the integration. This will produce a sine function in \eq{tot-amp5} leading to $\cos(Qd_c - \ell_i\pi+\pi)$ structures in the cross section for a $\ful$ $\sigma$ electron capture. This suggests that for the same angular momentum the $\pi$ electron capture cross section will produce 180$^o$ out-of-phase resonances compared to a $\sigma$ electron capture, as clearly seen between the HOMO and 6$h$ results in Fig.\,2(a). This indicates that electronic structural information can also be accessed spectroscopically by Ps formation from fullerenes.
\begin{figure}[h!]
\includegraphics[width=8cm]{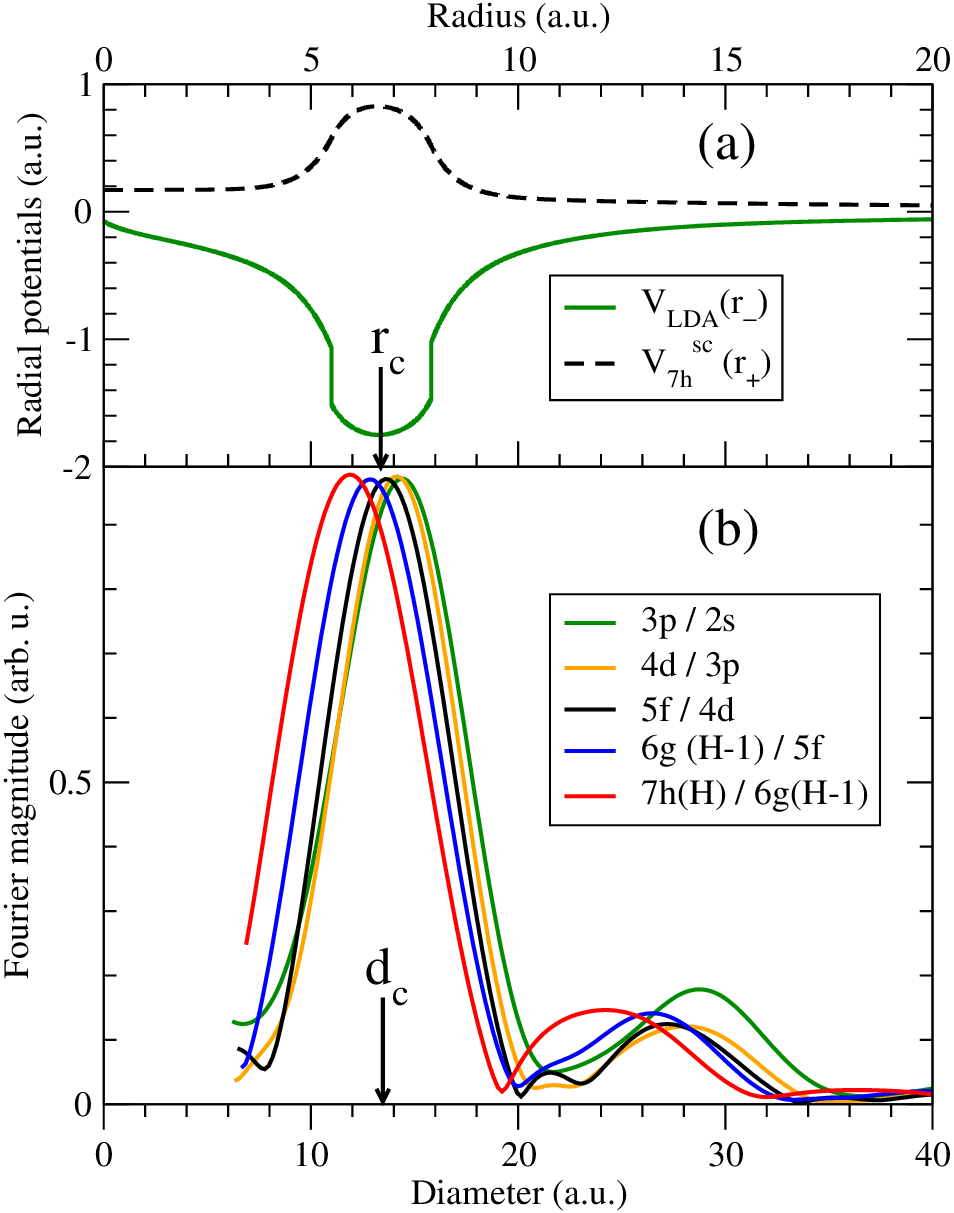}
	\caption{(Color online) (a) $\ful$ LDA radial potential and the positron scattering potential for the HOMO capture. (b) The Fourier transform magnitudes of Ps($1s$) cross section ratios for captures from various $\ful$ $\pi$ levels.}
	\label{figure3}
\end{figure}

An elegant way to bring out the connection of diffraction resonances with the fullerene diameter is to evaluate the Fourier spectra of the cross sections as a function of $Q$. To generate the input signals with only the resonances on a flat, non-decaying background we considered ratios of the results of two consecutive angular levels of $\pi$ electrons. Calculated Fourier magnitudes of these ratios in the reciprocal (radial) coordinate are displayed in Fig.\,3(b). All the curves exhibit strong peaks located close to the the diameter $d_c$, as expected. Note that the small, systematic offset of the peaks towards lower values with the increasing angular momentum is another signature of the fact that the continuum waves are Coulomb distorted and so are more complicated than simple plane waves used in the analysis. In fact, we anticipate this variation by noting in Fig.\,2(b) the separations $\Delta Q$ of 0.49 a.u.\ and 0.43 a.u.\ respectively for HOMO/HOMO-1 and 3p/2s ratio and then determining their reciprocal ($2\pi/\Delta Q$) values as 12.8 a.u.\ and 14.6 a.u.\ being somewhat different. In any case, these Fourier reciprocal spectra unequivocally supports the theme that the host of broad resonances are indeed the fringe patterns in the energy domain for a Ps formation channel where the forward Ps-emission diffracts in energy off the shell -- a spherical slit. Furthermore, our calculations (not shown) for the formation of excited Ps($2s$) produced similar resonance structures. The Ps($2s$) channel is attractive, since it can be monitored optically~\cite{murtagh2009}. Also, using the C$_{240}$ fullerene we found (not shown) the resonances to be more compact in energy than $\ful$, since C$_{240}$ is a larger diffractor.

In conclusion, we calculate the Ps formation cross section in the CDW-FS method for electron captures from various $\ful$ levels. The $\ful$ ground state structure is modeled by a simple but successful LDA frame. Hosts of strong and broad shape resonances in the Ps formation are predicted. The resonances are engendered from a diffraction effect in the Ps formation process localized on the fullerene shell. The effect should be universal for Ps formation from nanosystems, including metal clusters, carbon nanotubes, or even quantum dots that confine finite-sized electron gas. The work ushers a new research direction to apply Ps formation spectroscopy to gas-phase nanosystems which we hope can begin with fullerenes, since they currently enjoy significant attraction in precision measurements.

\begin{acknowledgments}
The research is supported by the National Science Foundation, USA.
\end{acknowledgments}


\begin{thebibliography}{10}

\bibitem{laricchia2008}
G. Laricchia, S. Armitage, \'{A}. K\"{o}v\'{e}r, and D.J. Murtagh, Ionizing collisions by positrons and positronium impact on the inert atoms,
Adv.\ At.\ Mol.\ Opt.\ Phys. \textbf{56}, 1 (2008).


\bibitem{schultz1988}
P.J. Schultz and K.G. Lynn, Interaction of positron beams with surfaces, thin films, and interfaces
\rmp\, \textbf{60}, 701 (1988).

\bibitem{green2015}
D.G. Green and G.F. Gribakin, γ-Ray spectra and enhancement factors for positron annihilation with core electrons, \prl\, \textbf{114}, 093201 (2015).

\bibitem{kauppila2004}
W.E. Kauppila, E.G. Miller, H.F.M. Mohamed, K. Pipinos, T.S. Stein, and E. Surdutovich, Investigations of positronium formation and destruction using 3γ/2γ-annihilation-ratio measurements, \prl\, \textbf{93}, 113401 (2004).


\bibitem{prantzos2011}
N. Prantzos, C. Boehm, A.M. Bykov, R. Diehl, K. Ferri\`{e}re, N. Guessoum, P. Jean, J. Knoedlseder, A. Marcowith, I.V. Moskalenko, A. Strong, and G. Weidenspointner, The 511 keV emission from positron annihilation in the Galaxy, \rmp\, \textbf{83}, 1001 (2011).

\bibitem{kavetskyy2014}
T. Kavetskyy, V. Tsmots, A. Kinomura, Y. Kobayashi, R. Suzuki, H.F.M. Mohamed, O. Sausa, V. Nuzhdin, V. Valeev, and A.L. Stepanov, Structural defects and positronium formation in 40 keV B$^+$-implanted polymethylmethacrylate, J.\ Phys.\ Chem.\ B \textbf{118}, 4194 (2014).

\bibitem{shu2016}
K. Shu, X. Fan, T. Yamazak, T. Namba, S. Asai, K. Yoshioka, and M. Kuwata-Gonokami, Study on cooling of positronium for Bose-Einstein condensation, J.\ Phys.\ B \textbf{49}, 104001 (2016).

\bibitem{mor2014}
O. Morandi, P.-A. Hervieux, and G. Manfredi, Bose-Einstein condensation of positronium in silica pores, Phys.\ Rev.\ A \textbf{89}, 033609 (2014).

\bibitem{cassidy2008}
D.B. Cassidy and A.P. Mills, Jr., Interactions Between Positronium Atoms in Porous Silica,
\prl\, \textbf{100}, 013401 (2008).

\bibitem{karshenboim2004}
S.G. Karshenboim, Precision study of positronium: testing bound state QED theory,
Int.\ J.\ Mod.\ Phys.\ A \textbf{19}, 3879 (2004).

\bibitem{cassidy2007}
D.B. Cassidy and A.P. Mills, Jr., The production of molecular positronium,
Nature \textbf{449}, 195 (2007).

\bibitem{mcconnell2016}
R. McConnell, G. Gabrielse, W.S. Kolthammer, P. Richerme, A. M\"{u}llers, J. Walz, D. Grzonka, M. Zielinski, D. Fitzakerley, and M.C. George, Large numbers of cold positronium atoms created in laser-selected Rydberg states using resonant charge exchange,
J.\ Phys.\ B \textbf{49}, 064002 (2016).

\bibitem{ferragut2010}
R. Ferragut, A. Calloni, A. Dupasquier, G. Consolati, F. Quasso, M.G. Giammarchi, D. Trezzi, W. Egger, L. Ravelli, M.P. Petkov, S.M. Jones, B. Wang, O.M. Yaghi, B. Jasinska, N. Chiodini, and A. Paleari, Positronium formation in porous materials for antihydrogen production,
J.\ Phys.\ Conf.\ Ser. \textbf{225}, 012007 (2010).

\bibitem{crivelli2014}
P. Crivelli, D.A. Cooke, and S. Friedreich, Experimental considerations for testing anti-matter gravity using positronium 1S-2S spectroscopy, Int.\ J. Mod.\ Phys.\ Conf. Ser.\ \textbf{30}, 1460257 (2014).

\bibitem{Perez2015}
P. P{\' e}rez, D. Banerjee, F. Biraben et al., The GBAR antimatter gravity experiment, Hyperfine Interact \textbf{233}, 21 (2015).

\bibitem{yamanaka2001}
N. Yamanaka and Y. Kino, Time-dependent coupled-channel calculations of positronium-formation cross sections in positron-hydrogen collisions,
\pra\, \textbf{64}, 042715 (2001).

\bibitem{kamali2001}
M.Z.M. Kamali and K. Ratnavelu, Positron-hydrogen scattering at low intermediate energies,
\pra\, \textbf{65}, 014702 (2001).

\bibitem{kar2000}
S. Kar and P. Mandal, Positronium formation in positron-hydrogen scattering using Schwingers principle,
\pra\, \textbf{62}, 052514 (2000).

\bibitem{sen2009}
S. Sen and P. Mandal, Positron-helium collisions: Positronium formation using the distorted-wave approximation,
\pra\, \textbf{80}, 062714 (2009).

\bibitem{ghanbari2013}
E. Ghanbari-Adivi and A.N. Velayati, Comparative study of the three- and four-body boundary-corrected Born approximations for positronium formation, J.\ Phys.\ B \textbf{46}, 065204 (2013).

\bibitem{mceachran2013}
R.P. McEachran and A.D, Stauffer, Positronium formation in the noble gases, J.\ Phys.\ B \textbf{46}, 075203 (2013).

\bibitem{gianturco1996}
F.A. Gianturco and R. Melissa, Positronium formation in positron-alkali-metal-atom collisions: An optical potential approach,
\pra\, \textbf{54}, 357 (1996).

\bibitem{pandey2016}
M.K. Pandey, Y-C. Lin, and Y.K. Ho, Positronium formation in collisions between positrons and alkali-metal atoms (Li, Na, K, Rb and Cs) in Debye plasma environments,
J.\ Phys.\ B \textbf{49}, 034007 (2016).

\bibitem{biswas2002}
P.K. Biswas, J.S.E. Germano, and T. Frederico, Positron-hydrogen molecule scattering considering the positronium-formation channel,
J.\ Phys.\ B \textbf{35}, L409 (2002).

\bibitem{sueoka2000}
O. Sueoka, M.K Kawada, and M Kimura, Total and positronium formation cross-sections in polyatomic molecules, Nuc.\ Instrum.\ Methods Phys.\ Res.\ B \textbf{171}, 96 (2000).

\bibitem{hervieux2006}
P.-A. Hervieux, O.A. Foj\'{o}n, C. Champion, R.D. Rivarola, and J. Hanssen, Positronium formation in collisions of fast positrons impacting on vapor water molecules, J.\ Phys.\ B \textbf{39}, 409 (2006).

\bibitem{garner1998}
A. Garner, A. \"{O}zen A, and G. Laricchia, Positronium beam scattering from atoms and molecules,
Nucl.\ Instrum.\ Methods Phys.\ Res.\ B \textbf{143}, 155 (1998).

\bibitem{machacek2016}
J.R. Machacek, F. Blanco, G. Garcia, S.J. Buckman, and J.P. Sullivan, Regularities in positronium formation for atoms and molecules,
J.\ Phys.\ B \textbf{49}, 064003 (2016).

\bibitem{eldrop1983}
M. Eldrup, A. Vehanen, P.J. Schultz, and K.G. Lynn, Positronium formation and diffusion in a molecular solid studied with variable-energy positrons, \prl\, \textbf{51}, 2007 (1983).

\bibitem{wang1998}
C.L. Wang, K. Hirata, J. Kawahara, and Y. Kobayashi, Electric-field dependence of positronium formation in liquids and polymers,
\prb\, \textbf{58}, 14864 (1998).

\bibitem{cabral-prieto2013}
A. Cabral-Prietoa, I. Garc\'{i}a-Sosaa, R. L\'{o}pez-Casta\~{n}aresb, and O. Olea-Cardosob, Positronium annihilation in LTA-type zeolite,
Micropor.\ Mesopor.\ Mater. \textbf{175}, 134 (2013).

\bibitem{cooper2016}
B.S. Cooper, A.M. Alonso, A. Deller, L. Liszkay, and D.B. Cassidy, Positronium production in cryogenic environments, \prb\, \textbf{93}, 125305 (2016).

\bibitem{crivelli2014mof}
P. Crivelli, D. Cooke, B. Barbiellini, B.L. Brown, J.I. Feldblyum, P. Guo, D.W. Gidley, L. Gerchow, and A.J. Matzger, Positronium emission spectra from self-assembled metal-organic frameworks, \prb\, \textbf{89}, 241103(R) (2014).

\bibitem{jones2015}
A.C.L. Jones, H.J. Goldman, Q. Zhai, P. Feng, H.W.K. Tom, and A.P. Mills, Monoenergetic Positronium Emission from Metal-Organic Framework Crystals, \prl\, \textbf{114}, 153201 (2015).

\bibitem{andersen2016}
S.L. Andersen, D.B Cassidy, J. Chevallier, B.S. Cooper, A. Deller, T.E. Wall, and U.I. Uggerhoj, Positronium emission and cooling in reflection and transmission from thin meso-structured silica films, J.\ Phys.\ B \textbf{49}, 204003 (2016).

\bibitem{fojon2001}
O. A. Foj\'{o}n, R. D. Rivarola, J. Hanssen, and P.A. Hervieux, Positronium formation in positron-simple metal cluster collisions,
J.\ Phys.\ B \textbf{34}, 4279 (2001).

\bibitem{jaenkaelae2011}
K.~J\"{a}nk\"{a}l\"{a}, M.~Tchaplyguine, M.~-H. Mikkel\"{a}, O.~Bj\"{o}meholm, and H.~Huttula, Photon energy dependent valence band response of metallic nanoparticles, \prl\, \textbf{107}, 183401 (2011).

\bibitem{ruedel2002oscExp}
A. R{\"u}del, R. Hentges, U. Becker, H.S. Chakraborty, M.E. Madjet, and J.M. Rost, Imaging delocalized electron clouds: photoionization of C$_{60}$ in Fourier reciprocal space, \prl\, \textbf{89}, 125503 (2002).

\bibitem{madjet2008}
M.~E. Madjet, H.~S. Chakraborty, J.~M. Rost, and S.~T. Manson, Photoionization of C$_{60}$: a model study,
\newblock J.\ Phys.\ B {\bf 41}, 105101 (2008).

\bibitem{ciappina2007}
M.~F. Ciappina, A.~Becker, and A.~Jaro\'{n}-Becker, Multislit interference patterns in high-order harmonic generation in C$_{60}$,
\newblock \pra\, {\bf 76}, 063406 (2007).

\bibitem{van1994exchange}
R. van Leeuwen and E.J. Baerends, Exchange-correlation potential with correct asymptotic behavior,
\pra\, \textbf{49}, 2421 (1994).

\bibitem{troullier1992}
N.~Troullier and J.~L. Martins, Structural and electronic properties of C$_{60}$,
\newblock \prb\, {\bf 46}, 1754 (1992).

\bibitem{weaver1991}
J.~H. Weaver, J.~L. Martins, T.~Komeda, Y.~Chen, T.~R. Ohno, G.~H. Kroll, and N.~Troullier, Electronic structure of solid C$_{60}$: experiment and theory,
\newblock \prl\, {\bf 66}, 1741 (1991).

\bibitem{vos1997}
M.~Vos, S.~A. Canney, I.~E. McCarthy, S.~Utteridge, M.~T. Michalewicz, and E.~Weigold, Electron-momentum spectroscopy of fullerene,
\newblock \prb\, {\bf 56}, 1309 (1997).

\bibitem{scully2005volumeExp}
S.W.J. Scully, E.D. Emmons, M.F. Gharaibeh, R.A. Phaneuf, A.L.D. Kilcoyne, A.S. Schlachter, S. Schippers, A. M{\"u}ller, H.S. Chakraborty, M.E. Madjet, and J.M. Rost, \prl\, \textbf{94}, 065503 (2005).

\bibitem{zamkov2004}
M. Zamkov, N. Woody, S. Bing, H.S. Chakraborty, Z. Chang, U. Thumm, and P. Richard, Time-Resolved Photoimaging of Image-Potential States in Carbon Nanotubes, \prl\, \textbf{93}, 156803 (2004).

\bibitem{fojon1996}
O.A. Foj\'{o}n, R.D. Rivarola, R. Gayet, J. Hanssen and P.-A. Hervieux, Continuum-distorted-wave–final-state approximation in positron-hydrogenic atom (ion) collisions with positronium formation, \pra\, \textbf{54}, 4923 (1996).

\bibitem{tang1992}
S. Tang and C.M. Surko, Angular dependence of positronium formation in molecular hydrogen, \pra\, \textbf{47}, R743 (1992).

\bibitem{shipman2015}
M. Shipman, S. Armitage, J. Beale, S.J. Brawley, S.E. Fayer, A.J. Garner, D.E. Leslie, P. Van Reeth, and G. Laricchia, Absolute Differential Positronium-Formation Cross Sections, \prl\, \textbf{115}, 033401 (2015).

\bibitem{murtagh2009}
D. J. Murtagh, D. A. Cooke, and G. Laricchia, Excited-State Positronium Formation from Helium, Argon, and Xenon, \prl\, \textbf{102}, 133202 (2009).

\end{thebibliography}

\end{document}